\documentstyle[11pt]{article}
\title{\bf
       INSTANTANEOUS CONFIGURATIONS OF THE BLOCH WALLS IN\\
          A TWO-DIMENSIONAL AND S=1/2 MODEL}
\author{
{\bf   Z.N\'eda$^{a,b}$ and G.Lipi$^b$  } \\
{\small\it
      $^a$Babes-B\'olyai University, Dept. of Physics}   \\
{\small\it
      str. Kogalniceanu 1, 3400 Cluj, Romania  }   \\
     \and
{\small\it and  }\\
{\small\it
       $^b$University of Bergen, Dept. of Physics   }\\
{\small\it
         All\'egaten 55, N-5007 Bergen, Norway}  }
\date{}

\begin{document}

\maketitle

\begin{center}

Abstract\\
\end{center}
We show that instantaneous configurations of 180$^o$ domain walls
constructed on a
square lattice in a two-dimensional and $S=1/2$ Ising-type
model exhibit fractal structure. The fractal dimension depends on the
coupling parameters and it is a continuous function of the temperature.
The wall thickness in the neighbourhood of $T_c$ presents scaling properties
in good agreement with the classical theory by Landau.

\vskip 40pt

\vfill \eject

\section*{1. Introduction}

   The theory for the average configuration of the Bloch walls is known
since the famous paper of Landau and Lifsitz \cite{1}. In this paper we
study the instantaneous configurations of the 180$^o$ Bloch walls for
$ S=1/2$ spins on a two-dimensional square lattice.
The hamiltonian of the problem is
considered of Ising-type \cite{2}. For square lattice, considering
interactions between  first-nearest neighbours only, the Ising model is
well understood
theoretically \cite{3}. If we introduce interactions with
next-nearest neighbours or domain
boundaries the problem becomes much more difficult and usually  studied
by computer simulation \cite{4,5,6}. This is the method we followed in the
present paper too. In the meantime the fractalness of the Ising
configurations near the critical temperature \cite{7,8}, or the fractal
structure of domains in a random Ising magnet \cite{9},
suggest that fractal theory
\cite{10,11} could be very adecvat to characterize  instantaneous
configurations of domain walls.
This problem, in the more general context of the roughening of interfaces
and their fractal character, was considered in several papers \cite{12,13,14}.
One can learn from these that the problem also presents interest in the
study of the diffusion front for interacting particles, invasion and corrosion.
Experiments \cite{15} show that similar structures appear for two-phase
liquid flow, where the interface of  two immiscible fluids exhibit
fractal structure dependent of the Reynolds number characteristic for the flow.

However computer simulations performed in \cite{12,13,14} for the diffusion
front of interacting particles or Ising models do not reveal the possibility
of variation for the fractal structure in correlation
with relevant physical parameters of the problem.
So in this context we proposed to study the Bloch wall structure obtained
by a computer model in function of temperature and characteristic interaction
parameters. In this way the problem could present interest  not only for
magnetism but in a more general view completing the basic ideas of
\cite{12,13,14}.

\section*{2. The Method}

   We constructed a heatbath dynamics model to simulate the dynamics of
$S=1/2$ spins on a square lattice, considering first-nearest
neighbour and next -nearest
 neighbour interactions between the spins. We imposed periodic boundary
conditions in the direction parallel to the wall, and mirror-symmetrical
boundary conditions in the direction perpendicular to the wall.
The size of the lattice were also varied in the interval 100-160 lattice sites
for the direction perpendicular to the wall (vertical direction), and 300-600
lattice sites for the direction parallel to the wall (horizontal direction).
At each step of
the simulation we choose spins randomly  , and apply the well-known
heatbath dynamics \cite{16}. In this dynamics the spin flip probability for
the spin $S_{i,j}$  is given by:
\begin{equation}
   W_{i,j}=\frac{exp( \beta E_{i,j} )}{exp( -\beta E_{i,j} ) +
exp( \beta E_{i,j})}.
\end{equation}
where $E_{i,j}=E_{i,j}^{o}+E_{i,j}^{1}$ , is the sum of the
interaction energies of
the considered spin with his first-nearest neighbours ,
$E_{i,j}^{o}$, and next-nearest neighbours ,$E_{i,j}^{1}$.
\begin{equation}
   E_{i,j}^{o} = -J_{o} \sum_{k} ( S_{i,j} S_{i+k,j} + S_{i,j} S_{i,j+k} ),
\end{equation}
\begin{equation}
   E_{i,j}^{1} = -J_{1} \sum_{k} ( S_{i,j} S_{i+k,j+k} +
 S_{i,j} S_{i+k,j-k} ).
\end{equation}
(Summations are for $k=+/-1$ only.)
To verify our model first we studied some single-domain
dynamics with first-nearest
 neighbour interactions. Starting the simulation from  completly random
or completly ordered configurations we compared our results
with theoretical ones.
So , we got for the critical temperature ,$T_c$, the value predicted by
the theory \cite{3} :
\begin{equation}
  T_c=\frac {1}{ 2 ln(1+\sqrt{2})}.\frac {J_o}{k}
   \approx 0.5673... \frac{J_o}{k}
\end{equation}
Starting the simulation from a perfect strait-line $180^o$ Bloch wall (Fig.1a)
we looked for the equilibrium shape of the domain boundaries.
We define the instantaneous domain boundary ( interface ) as the curve
separating the two inversly magnetized domains, and which is continuously
connected by first-nearest neighbours with the same spin orientation.
Every site of this curve must have at least one first-nearest neighbour with
inverse spin orientation, and the curve must realize a percolation in
horizontal direction.
This interface was detected by a special program after the equilibrium
configurations were reached. Characteristic result is presented in Fig.2 .

The thickness of the interface ,$\lambda$ , is considered as the number of rows
in which the average magnetization performs the transition between the
two $+m$ and $-m$ equilibrium values for the two domains. (This is $\lambda$
in Fig. 3).

Due to the complexity of these boundaries we used also the fractal theory
for their description. The fractal structure was studied by calculating
the fractal dimension ,D, with the simple box-counting method \cite{10}.
Working
on lattices with the mentioned sizes, we could apply our box-counting method
on the length-scale of 1-100 lattice constants. This interval is large enough
for evidencing fractal structures.

We considered the equilibrium configuration reached when the fractal dimension
and the wall-thickness of the interface had no monotonic variations, only
statistical fluctuations in time. This happend, depending on the considered
temperature between 700-10 000 iterations/spin. The obtained interfaces, after
this equilibrium was reached had complicate shapes and large fluctuations
in time, but in statistical sense  were stable ( Fig. 1 f,g).

\section*{3. Results}

	Characteristic time evolution of the structure is presented in Fig. 1.
Once the statistical equilibrium configuration reached we studied the
structure of the interface.
   The equilibrium shape of our domain walls had important structural
variations with temperature. This is illustrated in Fig. 4. for a model with
only first-nearest neighbour interactions.
As we mentioned earlier we studied two main aspects of the structure :
the fractal properties by the fractal dimension and the wall thickness.

\subsection*{a. Fractal dimension of the interface}

Considering only first-nearest neighbour interactions with $T_c=1000 K$ we
studied the variation of the fractal dimension ,$D$, with the temperature. The
results are plotted  in Fig.5 .
Including now next-nearest neighbour interactions too, with $T_o=560 K$ and
$T_1=110 K$ ($T_o=0.5673\frac{J_o}{k}$ , $T_1=0.5673\frac{J_1}{k}$ ,
$T_c=720 K$), we got the results presented in Fig. 6 .
As mentioned earlier the fractal dimension was calculated on the interval
of 1-100 lattice constants. Characteristic result for the fit is shown in
Fig. 7. In general for all the studied boundaries the fit quality was
good, indicating nice fractal structures on the whole interval of 1-100
lattice constants. The fit was made using at least 80 points, and the
worst value for the percent of residuals about mean explained for
fitting the log-log plot with a line in the box-counting method was 98\% !

Our results from Fig.5 and Fig. 6 suggest that at temperatures
small enough comparative to the
critical one, the instantaneous configuration of the considered Bloch walls
can be well described by a normal one-dimensional curve. For temperatures
between a value ,$T_f$, and $T_c$ the fractal dimension is,
in a good approximation, a linear function
of the temperature. This $T_f$ , characteristic temperature depends on the
chosen interaction parameters. So, the results qualitatively are not affected
by the interaction parameters, the only difference that appear is the shifting
of the characteristic temperature , $T_f$. Extrapolating our results for the
limit $T=T_c$, the interface tend to a structure with the $D=1.25$
fractal dimension for both of the considered models.

Identical results were found in \cite{15} ,by
studying the dynamical behaviour of the interphase between two immiscible
fluids flowing in a tube. A clear increase of the fractal dimension of the
interface as a function of total fluid velocity was found, and the fractal
dimension had the same kind of variation with the fluid velocity as those
obtained in Fig.5 and Fig.6 .

\subsection*{b. Wall thickness}

	For the equilibrium configurations the wall thickness exhibits the
variations plotted in Fig. 8. Considering temperatures close to the critical
one, we found that the variation could be well-described by the classical
theory , which predict:
\begin{equation}
\lambda \sim \sqrt{\frac{J}{K}}              ,
\end{equation}
(here J is the exchange constant and K the anisotropy constant) see for
example \cite{17}. The only source of the anisotropy in our lattice
is due to the preferenced spin orientation. In this sens,
one can consider for our model the anisotropy constant proportional with
the average magnetization of the domains.

For our system the magnetization ,$M$ ,scales like:
\begin{equation}
M \sim \sqrt {1-\frac{T}{T_c}},
\end{equation}
in the vicnity of the critical temperature ($T<T_c$). This fact is in
accordance with the theory \cite{3} which predict this scaling law for
every row with finite distance from the boundary.

So ,by applying equation (5) the theory predict:
\begin{equation}
\lambda \sim {(1-\frac{T}{T_c})}^{-\frac{1}{4}} .
\end{equation}
In Fig. 7 we plotted the variation of $\lambda$ in function of
$(1-\frac{T}{T_c})$. By fitting these with a power-law function we
found for the scaling exponent the value $-0.256$ for simulations with
first-nearest neighbour interactions only, and $-0.268$ for the case
when we included next-nearest neighbour interactions too.

With the assumption made, that for our model one should consider the
anisotropy constant proportional with the average magnetization,
we got that our results are in good agreement with the classical theory for
the Bloch wall thickness.

\section*{4.Conclusions}

    We obtained in  our two-dimensional model, that the instantaneous
configurations of the $180^o$ Bloch walls for $S=1/2$ spins exhibit fractal
structure in the $T_f < T < T_c$ temperature interval. In this region the
fractal dimension is a linear function of the temperature. The fractal
dimension
is in a strong correlation with relevant physical quantities of the considered
problem, it is very adecvat for characterizing the considered domain walls,
and for detecting the equilibrium configuration in computer simulations.
The wall thickness presents a power-law scaling in the neighbourhood of
$T_c$ ($T<T_c$) in accordance with classical theories.
The observed connection with geometrical aspects of  two-phase liquid flow is
interesting, and suggests, universalities in the geometry of
two-phase boundaries.
This problem, of the universality or not, could be interesting and suggests
further studies.

\section*{Acknowledgements}

\samepage

   This work was done during a scholarship offered by the Physics Department
of the University of Bergen.
So, we are grateful to the staff of the Theoretical-
and Reservoire- Physics groups, and especially for the generous help from
L. Csernai. Discussions with J. P. Hansen are gratefully acknowledged.
During our visit at the Energy Technology Institute in Oslo we had interesting
discussions with the team of  A. Skjeltorp. We also thank to
Y. Brechet of ENSEEG-LTPCM Grenoble,  M. Coldea of the Babes-Bolyai
University of Cluj, and to  J. Kertesz of BME-Budapest for their
continuous help they given to us in elaborating this paper.

\newpage
\section*{Figure Captions}

\vspace{.25in}

Fig.1.  Time evolution of a domain wall after 0, 1, 8, 50,
100, 1000 and 10000 simulation steps per pixel (A, B, C,...G
respetively).
\vspace{.25in}

Fig.2.   Detection of the domain wall.
\vspace{.25in}

Fig.3.   Typical variation for average spin per row , $<S>$, through the wall.
         $\lambda$ is the wall thickness.
\vspace{.25in}

Fig.4.   Equilibrium shapes for the  instantaneous
configuration of the domain wall
at different temperatures ( the temperatures are in K ).
The critical temperature is $T_c=1000 K$.
\vspace{.25in}

Fig.5.  Variation of the fractal dimension of the domain wall with
temperature for
simulations considering first-nearest neighbour interactions only
and $T_c=1000 K$ ( the temperatures are in K ).
\vspace{.25in}

Fig.6.  Variation of the fractal dimension of the domain wall
with temperature for simulations considering first- and next-nearest neighbour
interactions. $T_o=560 K$ , $T_1=110 K$, $T_c=720K$
($T_o=0.5673 \frac{J_o}{k}$ , $T_1=0.5673 \frac{J_1}{k}$ ).
\vspace{.25in}

Fig.7.  Characteristic result for the fit applying the box-counting
        method. Here $L$ is the linear size of the box ( the units for
        $L$ are lattice constants ) and $N(L)$ represents the
        number of boxes necessary to cover our structure with boxes of size
        $L$. The slope
        of the best-fit line is the fractal dimension of the considered
        interface.
\vspace{.25in}

Fig.8.  Variation of the wall thicness, $\lambda$, ( the units are in
        lattice constants)  in function of
        $T-T_c$ ( $T$ the temperature, and $T_c$ the critical temperature).
        The "first model" are results for considering first-nearest neighbour
        interactions only, the "second model" are results considering
        next-nearest neighbour interactions too.
        In the neighbourhood of $T_c$ the datas were fit by :
         $\lambda=103\cdot(T-T_c)^{-0.256}$ for the first model and
         $\lambda=126\cdot(T-T_c)^{-0.268}$ for the second one.
\vspace{.25in}

\end{document}